\documentclass{aa}

\usepackage{graphicx}
\usepackage[dvipsnames]{color}
\usepackage{natbib}

\begin{document}

\newcommand{\bl}[1]{\mbox{\boldmath$ #1 $}}

\title{The effect of external environment on the evolution of protostellar disks}
\titlerunning{The effect of external environment}

\author{Eduard I. Vorobyov \inst{1,2}, D. N. C. Lin \inst{3}, and Manuel Guedel \inst{1}}
\authorrunning{Vorobyov et al.}

\institute{University of Vienna, Department of Astrophysics,  Vienna, 1180, Austria; 
\email{eduard.vorobiev@univie.ac.at}
\and
Research Institute of Physics, Southern Federal University, Stachki Ave. 194, Rostov-on-Don, 
344090 Russia; 
\and
UCO/Lick Observatory, University of California, Santa Cruz, CA 95064, USA
}

\abstract
{}
{Using numerical hydrodynamics simulations we studied the gravitational collapse 
of pre-stellar cores of sub-solar mass embedded into a low-density external environment. }
{Four models with different magnitude and direction of rotation of the external environment with
respect to the central core were studied and compared with an isolated model. }
{ We found that the 
infall of matter from the external environment can significantly alter the disk properties as compared
to those seen in the isolated model. Depending on the magnitude and direction of 
rotation of the external environment, a variety of disks can form including compact ($\le 200$~AU) ones
shrinking in size due to infall of external matter with low angular momentum, as well as extended 
disks forming due to infall of external matter with high angular momentum. The former are usually stable
against gravitational fragmentation, while the latter are prone to fragmentation and formation 
of stellar systems with sub-stellar/very-low-mass companions. 
In the case of counterrotating external environment, very compact ($<5$~AU) and short-lived 
($\la \mathrm{a~few}~10^5$~yr) disks can form when infalling material has low angular momentum. 
The most interesting case is found for the infall of counterrotating external material 
with high angular momentum,
leading to the formation of counterrotating inner and outer disks separated by a deep gap 
at a few tens AU. The gap migrates inward due to accretion of the inner disk onto the protostar, 
turns into a central hole, and finally disappears giving way to the outer strongly 
gravitationally unstable disk. This model may lead to the emergence 
of a transient stellar system with sub-stellar/very-low-mass components counterrotating 
with respect to that of the star.
}
{}

\maketitle

\section{Introduction}
\label{intro}
Solar-type stars form from the gravitational collapse of dense molecular cloud cores, 
which in turn can be the product of supersonic turbulent compression \citep[e.g.][]{Padoan2001} or
slow expulsion of magnetic field due to ambipolar diffusion \citep[e.g.][]{Shu1987,BM1994}.
A large fraction of the collapsing core passes through a protostellar disk, formed due to 
conservation of the net angular momentum of the core, before finally landing on
a protostar.  The main accretion phase lasts until most of the parental core is accreted 
onto the disk plus protostar system or dispersed via feedback from stellar irradiation
and outflows. 
In this ''classic'' scenario, the core is usually described by an isolated, initially unstable 
Bonnor-Ebert-type sphere or perturbed singular isothermal sphere.

However, recent observations indicate that dense molecular cores constitute part of a complicated filamentary
structure in the spatial and velocity space, with individual filaments resembling a network 
of twisted and intersecting threads \citep{Hacar2013}. A general agreement between the measured 
core masses and those predicted from the linear stability analysis 
suggests that the cores are likely to form from the gravitational fragmentation of these filaments 
\citep{Hacar2011}. In this picture, prestellar cores may not be described by a perfectly spherical
shape and may be submerged into a varying density environment characterized by gas flows with
varying angular momentum. As a net result, it is likely that the protostar plus disk system 
may continue accreting from the external environment even after the parental core has completely dissipated.

This conjecture seems to be supported by recent observations showing that large star-forming
regions contain a population of relatively old stars (with age~$>$~10~Myr), 
which still have signatures of accretion disks \citep{Beccari2010,DeMarchi2013a,DeMarchi2013b}. 
\citet{Scicluna2014} 
explain this phenomenon by invoking Bondi-Hoyle accretion to rebuild a new disk around these stars 
during passage through a clumpy molecular cloud. While it is difficult to determine the proper 
stellar age of accreting stellar objects \citep{Baraffe2012} and
it is not clear if Bondi-Hoyle accretion is the right mechanism to describe accretion onto the disk, the fact that there may be quite old disks (with age much greater than the canonical
disk-dispersal $e$-folding time of 3~Myr) supports the idea 
of prolonged accretion.

Theoretical and numerical studies of clustered star formation also suggest that
accretion may continue from larger (than the initial extent of the parental core) 
spatial scales until the gas reservoir within the cluster is completely exhausted, leading to a
constant accretion time that is proportional to the global freefall time of the cluster 
\citep[e.g.][]{McKee2010,DunhamPPVI}.  In particular, numerical simulations 
of, e.g., \citet{Bate2010}, \citet{Fielding2014}, and \citet{Padoan2014} demonstrate that 
molecular clouds are turbulent and inherently chaotic environments, within which individual 
collapsing cores may accrete mass and angular at a highly non-steady rate from a network 
of filaments interconnecting the cores. It is not impossible that the angular momentum 
vector of the accreted material can undergo significant changes both in magnitude and direction. 

Finally, observations seem to indicate longer lifetimes of the
main accretion phase, 0.42~Myr \citep{Evans09}, than those derived from numerical simulations of 
{\it isolated} gravitationally unstable cores, 0.12~Myr \citep{Vorobyov2010,Dunham2012}, again implying
that mass infall onto the disk/star system may continue after the initial parental core has dissipated.

In this paper, we explore the effect of prolonged gas infall onto the properties 
of circumstellar disks. In particular, we consider prestellar cores submerged into a low-density, external environment with varying angular momentum. As a first approximation, we assume that the external material
has constant density and angular velocity.
More realistic initial conditions taken from numerical hydrodynamics simulations of clustered star formation
will be considered in a follow-up study.

\section{Model description}
\label{model}
Our numerical hydrodynamics model for the formation and evolution of a star plus disk system 
in the thin-disk limit is described in detail in 
\citet{VB10} and \citet{Vorobyov2013} and is briefly reviewed below for the reader's convenience.
We start our numerical simulations from the gravitational collapse of a {\it starless} cloud core
submerged into a low-density external environment, 
continue into the embedded phase of star formation, during which
a star, disk, and envelope are formed, and terminate our simulations after approximately 0.5 Myr of
disk evolution.
The protostellar disk occupies the inner part of the numerical polar grid
and is exposed to intense mass loading from the infalling envelope and external environment.

To avoid too small time steps, we introduce a ``sink cell'' at $r_{\rm sc}=5$~AU and 
impose a free inflow inner boundary condition
and a free outflow outer boundary condition so that that the matter is allowed to flow out of 
the computational domain but is prevented from flowing in. 
The sink cell is dynamically inactive; it contributes only to the total gravitational 
potential and secures a smooth behaviour of the gravity force down to the stellar surface.
During the early stages of the core collapse, we monitor the gas surface density in 
the sink cell and when its value exceeds a critical value for the transition from 
isothermal to adiabatic evolution, we introduce a central point-mass object.
In the subsequent evolution, 90\% of the gas that crosses the inner boundary 
is assumed to land on the central object. 
The other 10\% of the accreted gas is assumed to be carried away with protostellar jets.

The basic equations of mass, momentum, and energy transport in the thin-disk limit are
\begin{equation}
\label{cont}
\frac{{\partial \Sigma }}{{\partial t}} =  - \nabla_p  \cdot 
\left( \Sigma \bl{v}_p \right),  
\end{equation}
\begin{eqnarray}
\label{mom}
\frac{\partial}{\partial t} \left( \Sigma \bl{v}_p \right) &+& \left[ \nabla \cdot \left( \Sigma \bl{v_p}
\otimes \bl{v}_p \right) \right]_p =   - \nabla_p {\cal P}  + \Sigma \, \bl{g}_p + \\ \nonumber
& + & (\nabla \cdot \mathbf{\Pi})_p, 
\label{energ}
\end{eqnarray}
\begin{equation}
\frac{\partial e}{\partial t} +\nabla_p \cdot \left( e \bl{v}_p \right) = -{\cal P} 
(\nabla_p \cdot \bl{v}_{p}) -\Lambda +\Gamma + 
\left(\nabla \bl{v}\right)_{pp^\prime}:\Pi_{pp^\prime}, 
\end{equation}
where subscripts $p$ and $p^\prime$ refer to the planar components $(r,\phi)$ 
in polar coordinates, $\Sigma$ is the mass surface density, $e$ is the internal energy per 
surface area, 
${\cal P}$ is the vertically integrated gas pressure calculated via the ideal equation of state 
as ${\cal P}=(\gamma-1) e$,
$Z$ is the radially and azimuthally varying vertical scale height
determined in each computational cell using an assumption of local hydrostatic equilibrium,
$\bl{v}_{p}=v_r \hat{\bl r}+ v_\phi \hat{\bl \phi}$ is the velocity in the
disk plane, and $\nabla_p=\hat{\bl r} \partial / \partial r + \hat{\bl \phi} r^{-1} 
\partial / \partial \phi $ is the gradient along the planar coordinates of the disk. 
The dependence of $\gamma$ on temperature is parameterized based on the work of \citet{MI2000} with
$\gamma=7/5$ for the gas temperature less than 100~K and $\gamma=5/3$ otherwise\footnote{In 
the case that the gas temperature exceeds 2000~K, $\gamma$ is set to 1.1, but a limited numerical
resolution does not allow us to follow the second collapse of hot fragments}.
The gravitational acceleration in the disk plane, $\bl{g}_{p}=g_r \hat{\bl r} +g_\phi \hat{\bl \phi}$, takes into account self-gravity of the disk, found by solving for the Poisson integral 
\citep[see details in][]{VB10}, and the gravity of the central protostar when formed. 
Turbulent viscosity is taken into account via the viscous stress tensor 
$\mathbf{\Pi}$, the expression for which is provided in \citet{VB10}.
We parameterize the magnitude of kinematic viscosity $\nu$ using the $\alpha$-prescription 
with a spatially and temporally uniform $\alpha=1.0\times 10^{-2}$.

The radiative cooling $\Lambda$ in equation~(\ref{energ}) is determined using the diffusion
approximation of the vertical radiation transport in a one-zone model of the vertical disc 
structure \citep{Johnson03}
\begin{equation}
\Lambda={\cal F}_{\rm c}\sigma\, T_{\rm mp}^4 \frac{\tau}{1+\tau^2},
\end{equation}
where $\sigma$ is the Stefan-Boltzmann constant, $T_{\rm mp}={\cal P} \mu / R \Sigma$ is 
the midplane temperature of gas\footnote{This definition of the midplane temperature is accurate within
a factor of unity \citep{Zhu2012}}, $\mu=2.33$ is the mean molecular weight, $R$ is the universal 
gas constant, and ${\cal F}_{\rm c}=2+20\tan^{-1}(\tau)/(3\pi)$ is a function that 
secures a correct transition between the optically thick and optically thin regimes.  
We use frequency-integrated opacities of \citet{Bell94}.
The heating function is expressed as
\begin{equation}
\Gamma={\cal F}_{\rm c}\sigma\, T_{\rm irr}^4 \frac{\tau}{1+\tau^2},
\end{equation}
where $T_{\rm irr}$ is the irradiation temperature at the disc surface 
determined by the stellar and background black-body irradiation as
\begin{equation}
T_{\rm irr}^4=T_{\rm bg}^4+\frac{F_{\rm irr}(r)}{\sigma},
\label{fluxCS}
\end{equation}
where $T_{\rm bg}$ is the uniform background temperature (in our model set to the 
initial temperature of the natal cloud core)
and $F_{\rm irr}(r)$ is the radiation flux (energy per unit time per unit surface area) 
absorbed by the disc surface at radial distance 
$r$ from the central star. The latter quantity is calculated as 
\begin{equation}
F_{\rm irr}(r)= \frac{L_\ast}{4\pi r^2} \cos{\gamma_{\rm irr}},
\label{fluxF}
\end{equation}
where $\gamma_{\rm irr}$ is the incidence angle of 
radiation arriving at the disc surface (with respect to the normal) at radial distance $r$.

The stellar luminosity $L_\ast$ is the sum of the accretion luminosity $L_{\rm \ast,accr}=G M_\ast \dot{M}/2
R_\ast$ arising from the gravitational energy of accreted gas and
the photospheric luminosity $L_{\rm \ast,ph}$ due to gravitational compression and deuterium burning
in the stellar interior. The stellar mass $M_\ast$ and accretion rate onto the star $\dot{M}$
are determined using the amount of gas passing through
the sink cell. The properties of the forming protostar are 
calculated using a stellar evolution code described in \citet{Baraffe2010}.
This code is coupled with the main hydrodynamical code in real time. 
However, due to heavy computational load  the stellar evolution code is 
invoked to update the properties of the protostar 
only every 20~yr, while the hydrodynamical time step may be as small
as a few weeks and the entire duration of numerical simulations may exceed 1.0 Myr.
The input parameter for the stellar evolution code is the accretion rate 
$\dot{M}$, while the output are the radius and photospheric luminosity of the protostar. 
This coupling of the disk and protostar evolution allows for a self-consistent determination
of the radiative input of the protostar into the disk thermal balance, which is important for
the accurate study of disk instability and fragmentation.

\begin{figure}
  \resizebox{\hsize}{!}{\includegraphics{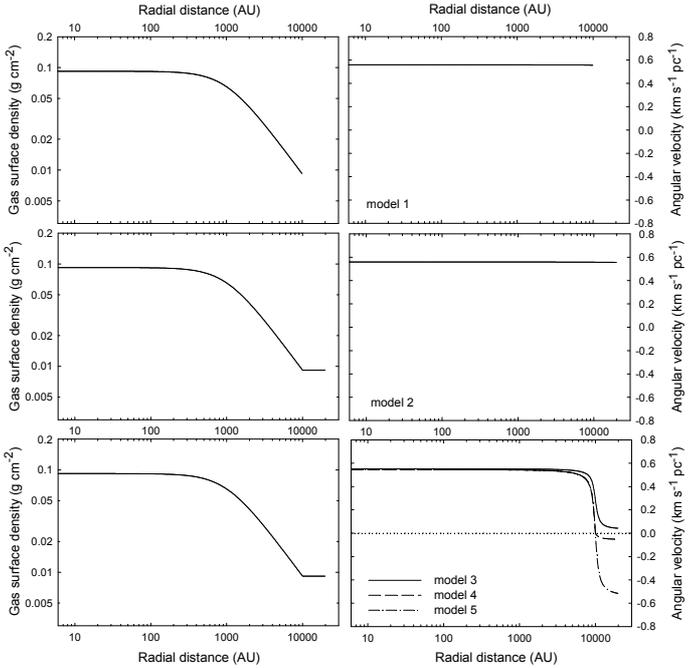}}
  \caption{Initial distribution of the gas surface density (left panels) and angular velocity (right
  panels) in five considered models. The horizontal dotted line marks the zero angular velocity.}
  \label{fig1}
  \hskip 0.1 cm
\end{figure}

\begin{table*}
\center
\caption{Model parameters}
\label{table1}
\begin{tabular}{cccccccccc}
\hline\hline
Model & $M_{\rm core}$ & $\beta_{\rm core}$ & $M_{\rm ext}$ &  $\beta_{\rm ext}$ & $\Omega_{\rm core}$  & $C$  & $r_{\rm 0}$ & $\Sigma_0$  & $R_{\rm core}$  \\
 & $M_\odot$ & $\%$ & $M_\odot$ & & km~s$^{-1}$~pc$^{-1}$ & & AU & g~cm$^{-2}$ & pc \\
\hline
1 & 0.63 & 0.53 & -- & -- & 0.56 & 0 & 1030 & $9.2\times 10^{-2}$ & 0.05 \\
2 & 0.63 & 0.65 & 1.0 & 2.5  & 0.56 & 1.0 & 1030 & $9.2\times 10^{-2}$ & 0.05 \\
3 & 0.63 & 0.48 & 1.0 & 0.033 & 0.56 & 0.1 & 1030 & $9.2\times 10^{-2}$ & 0.05 \\
4 & 0.63 & 0.36 & 1.0 & 0.019 & 0.56 & -0.1 & 1030 &  $9.2\times 10^{-2}$ & 0.05 \\
5 & 0.63 & 0.36 & 1.0 & 1.2  & 0.56 & -1.0 & 1030 & $9.2\times 10^{-2}$ & 0.05 \\
\hline
\end{tabular}
\end{table*}

\section{Initial conditions}
As the initial setup, we take a pre-stellar core submerged into a constant-density
external environment. For the initial surface density profile of the core, we adopt a simplified
form of a vertically integrated Bonnor-Ebert sphere \citep{Dapp09}
The resulting initial distribution of the gas surface density takes the following form
\begin{equation}
\label{dens}
\Sigma = \left\{
\begin{array}{ll}
{r_0 \Sigma_0 \over \sqrt{r^2+r_0^2}} \,\,\,\,\, 
       \mathrm{for}~r\le R_{\rm core}, \nonumber \\
\Sigma_{\rm ext} \hskip 0.8cm \mathrm{otherwise},            
\end{array} \right. 
\end{equation}
where $\Sigma_0$ is the gas surface
density at the center of the core, $r_0 =\sqrt{A} c_{\rm s}^2/\pi G \Sigma_0 $
is the radius of the central plateau of the core, $R_{\rm core}$ is the radius of the core, 
$c_{\rm s}$ is the initial sound speed,
and $\Sigma_{\rm ext}$ is the density of the external environment, the value of which
is set equal to the gas surface density at the outer edge of the core ($\Sigma_{\rm ext}=r_0 \Sigma_0/\sqrt{R_{\rm
core}^2+r_0^2}$). 
In all models the value of $A$ is set to 1.2 and the initial temperature is set to 10~K.

To study the effect of infall from the external environment characterized by 
varying angular momentum, we adopt the following form for
the initial radial profile of angular velocity $\Omega$ 
\begin{equation}
\label{angular}
\Omega = \left\{
\begin{array}{ll}
\Omega_{\rm core}  \hskip 0.9cm \mathrm{for}~r\ll R_{\rm core}, \nonumber \\
C \Omega_{\rm core}  \hskip 0.4cm \mathrm{for}~r\gg R_{\rm core},            
\end{array} \right. 
\end{equation}
where  $\Omega_{\rm core}$ is the angular velocity of the core and  
$C$ is a dimensionless scaling factor defining the angular velocity in the external environment.
The angular velocity profile between the core and the external environment is smoothly joined 
at the outer edge of the core.

We have considered 5 models, the initial setup for which is shown in Figure~\ref{fig1}.
In particular, the left panels present the radial distribution of $\Sigma$, while the right panels
show $\Omega$ as a function of radial distance. Model~1 represents a uniformly rotating isolated core
(no external environment) with mass $0.63~M_\odot$ and ratio of rotational to gravitational energy 
$\beta=0.51\%$. Models~2--5 consider the same core submerged into the external environment with 
different angular velocities. In particular, model~2 has a corotating external environment with 
the same angular velocity as that of the core, while in model~3 the angular velocity of 
the external environment is 10 times 
smaller than that of the core. Models~4 and 5 are characterized by a counterrotating external environment.
In particular, in model~4 the external angular velocity is ten times smaller by absolute value than
that of the core, while in model~5 the external angular velocity is equal to that of the core
by absolute value. In models~2-5 the radial distribution of gas surface density is identical.
The parameters of every model are provided in Table~\ref{table1}.

\begin{figure}
 \centering
  \resizebox{\hsize}{!}{\includegraphics{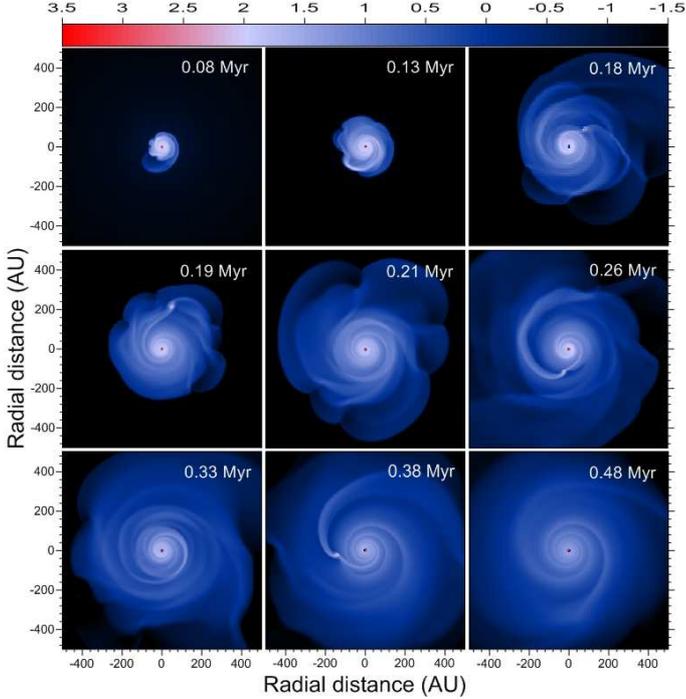}}
  \caption{Gas surface density maps in model~1 in the inner $1000\times1000$~AU box obtained
  at several evolution times (as marked in each panel) after the formation of the central protostar.
  The scale bar is in log g~cm$^{-2}$.}
  \label{fig2}
\end{figure}

\section{Results}

In this section, we start with considering the properties of a protostellar disk formed
from the gravitational collapse of an isolated pre-stellar core and continue by considering
models that take into account the effect of external environment.

\subsection{The isolated model}
As a prototype model, we consider an isolated prestellar core with mass $0.63~M_\odot$ and
ratio of rotational to gravitational energy $\beta=5.3\times10^{-3}$ (hereafter, model~1). 
This choice is motivated  by our previous numerical hydrodynamics simulations indicating 
that protostellar disks formed
from cores with similar characteristics are expected to be gravitationally unstable and prone to
fragmentation in the early embedded stage of stellar evolution \citep{Vor2011}. Figure~\ref{fig2} presents
the time evolution of  the gas surface density in model~1 in the inner $1000\times1000$~AU box. 
The whole computational region extends to 10000~AU. The time elapsed from the formation 
of the central star is indicated in each panel. The scale bar is in log~g~cm$^{-2}$ and 
the lower surface density threshold is set 
to $\log \Sigma=-1.5$~g~cm$^{-2}$ so that the black region is not empty but filled with a low density
material falling onto the disk from the parental core.

The protostellar disk in model~1 becomes gravitationally unstable at the very early stage 
of evolution (as manifested by a spiral structure already present at $t=0.08$~Myr) due 
to a continuing mass loading from the infalling core. 
The first fragment forms in the disk after $t=0.15$~Myr, when the disk mass and 
radius begin to exceed 0.1~$M_\odot$ and 100~AU, respectively. This is in agreement with many previous numerical and 
theoretical studies, indicating that the disk needs to acquire a critical mass ($\ga 0.1~M_\odot$) 
and grow to a critical size ($\ga 50-100$~AU) before gravitational fragmentation can ensue 
\citep{Stamatellos2008,Rice2010,Vor2011,Meru2012}. We used the fragment tracking algorithm described
in \citet{Vor2013} to calculate the masses of the fragments. The fragments were 
identified based on two conditions: 1) the fragment must be pressure-supported, with a negative
pressure gradient with respect to the center of the fragment and 2) the fragment must be 
kept together by gravity, with the potential well being deepest at the center 
of the fragment. The resulting masses of the fragments lie in the 2.5--5.0~$M_{\rm Jup}$ range.

One can see in Figure~\ref{fig2} that the process of disk fragmentation in model~1 
is not continuous but intermittent --
time intervals showing fragments are alternated by intervals showing no signs of disk 
fragmentation. The timescale with which the disk fragments is determined
by the ratio \citep{VZD2013} 
\begin{equation}
T_{\rm fr}={M_{\rm d} \over \dot{M}_{\rm infall} }, 
\label{Tfrag}
\end{equation} 
where $M_{\rm d}$ is the disk mass and 
$\dot{M}_{\rm infall}$ is the rate of mass infall from the parental core onto the disk. 
Taking $0.15~M_\odot$ and $10^{-6}~M_\odot$~yr$^{-1}$ for the typical values of $M_{\rm d}$ and $\dot{M}_{\rm
infall}$ (see Figs.~\ref{fig10} and \ref{fig11}), we obtain $T_{\rm fr}=0.15$~Myr, which 
appears to agree (within a factor of unity) with a visually estimated value of 
$T_{\rm fr}\approx 0.1$~Myr (see Fig.~\ref{fig2}). The lifetime of fragments is usually 
shorter than 0.1~Myr, either due to migration on the central protostar or destruction via 
tidal torques  \citep{Stamatellos2011, VZD2013}, 
resulting in prolonged periods of disk evolution without fragments.
Finally, it is worth noting that the disk forms  only one fragment at a time,
which implies that ejection of fragments  into the intercluster medium due to gravitational many-body
interaction in such a system is impossible \citep{BV2012}.

As many numerical studies indicate, a fragment can be driven into the disk inner 
regions and probably onto the star 
\citep{VB06,VB10,Machida2011,Cha2011}, ejected into the intracluster medium \citep{BV2012}, dispersed
by tidal torques \citep{Boley2010,Zhu2012} or even survive and settle onto quasi-stable, 
wide-separation orbits \citep{Vor2013}. In model~1, there are no accretion bursts 
that could be triggered by fragments migrating onto the star (see Fig.~\ref{fig11}). 
There are also no ejections or survival of the fragments. This leaves us only with
one outcome -- the fragments in model~1 have been tidally dispersed.

\subsection{The effect of external environment: corotating infall}
\begin{figure}
 \centering
   \resizebox{\hsize}{!}{\includegraphics{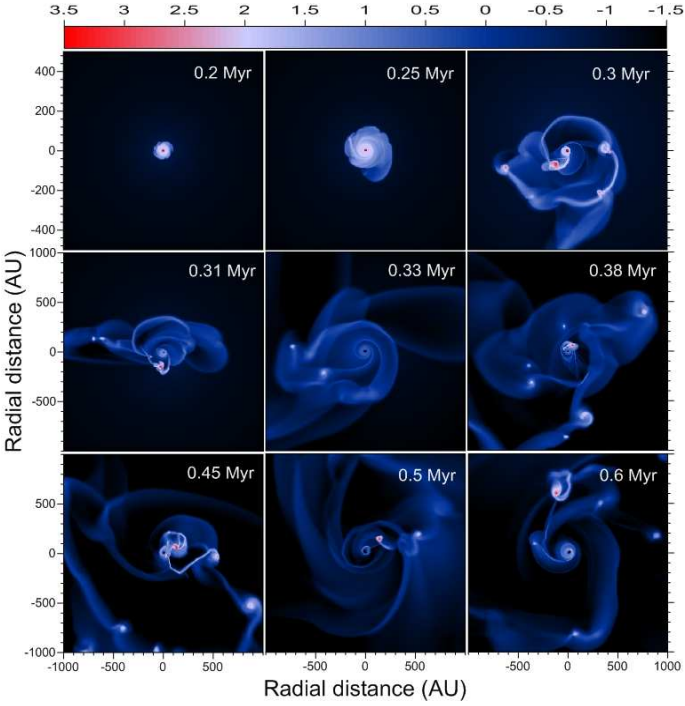}}
  \caption{Gas surface density maps in model~2 in the inner $1000\times1000$~AU box (top row) and 
  $2000\times2000$~AU box (middle and bottom rows) obtained
  at several evolution times (as marked in each panel) after the formation of the central protostar.
  The scale bar is in log g~cm$^{-2}$.}
  \label{fig3}
\end{figure}

In this section we consider models 2 and 3 in which a prestellar core is submerged 
in an external environment corotating with the core. 
Figure~\ref{fig3} presents the gas surface density maps in model~2 at the same evolution times as 
in Figure~\ref{fig2} for the isolated core. The top panels have a size of $1000\times 1000$~AU 
centered on the star, while the other panels have a twice larger size. 
The early evolution in model~2 ($t\le 0.12$~Myr) is similar to that in model~1---
the disk is gravitationally unstable showing a flocculent spiral structure and no signs of
fragmentation. This is explained by the fact that during this early period 
the disk accretes the matter from the parental core rather than 
from the external environment. 

The effect of the external environment becomes evident at
$t>0.15$~Myr when its matter starts reaching the disk surface. In this phase, the 
disk evolution in model~2 becomes strikingly different from that in the isolated model~1.  
Violent gravitational instability and fragmentation leads
to almost complete disintegration of the disk. The system now resembles a multi-component 
stellar system connected with dense filaments, within which 
individual components tend to form their own mini-disks. The masses of the fragments span
a range from a few Jupiters to very-low-mass stars, with the most massive fragment having a mass
of $0.18~M_\odot$.

\begin{figure}
  \resizebox{\hsize}{!}{\includegraphics{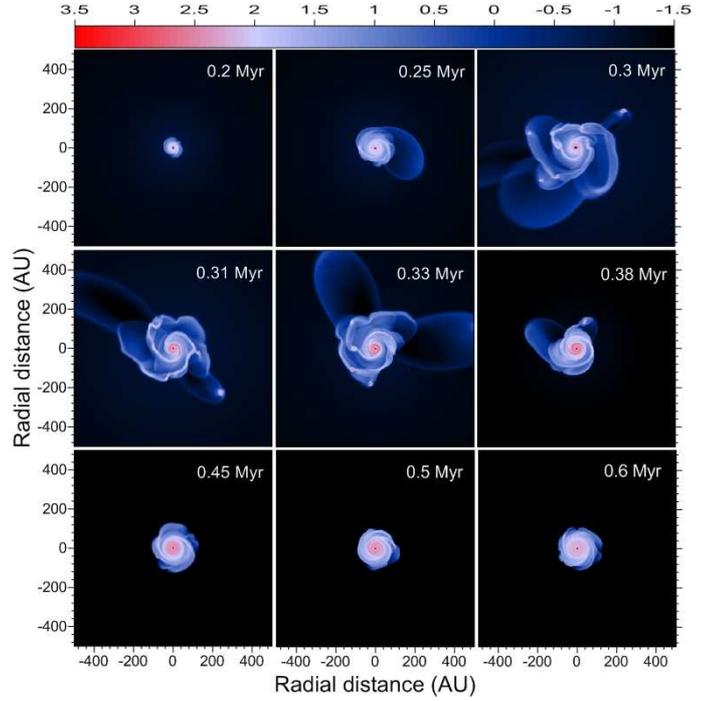}}
  \caption{Similar to Figure~\ref{fig2} but for model~3.}
  \label{fig4}
\end{figure}

As the next step, we consider model~3 characterized by the external environment rotating at 
a ten times smaller angular velocity than that of the core. 
Figure~\ref{fig4} presents the gas surface density maps in model~3 in the same spatial box 
of $1000\times1000$~AU and 
at similar evolution times as for the isolated model~1 (see Fig.~\ref{fig2}).
The early evolution at $t\le0.12$~Myr is similar in both models~1 and 3. The disk is weakly gravitationally
unstable and develops a flocculent spiral structure. A significant 
difference in the disk appearance between models 1 and 3 is seen after $t=0.15$~Myr, when infall from
the external environment triggers a vigourous gravitational instability and fragmentation in the disk.
Unlike model~1 however, the process of disk fragmentation in model~3 seems to be continuous 
(in the time interval of $t=0.15-0.25$~Myr) and is characterized by the presence of multiple fragments.
This is explained by the fact that the time period between disk fragmentation episodes $T_{\rm fr}=5\times
10^{4}$~yr, calculated using
equation~(\ref{Tfrag}) and typical disk masses and infall rates ($M_{\rm d}=0.25~M_\odot$ and 
$\dot{M}_{\rm infall}=5\times 10^{-6}~M_\odot$~yr$^{-1}$),
appears to be comparable to the longest migration times of fragments onto the central 
protostar \citep{VZD2013}. As a net result, some of the fragments are always present in the disk.
The masses of the fragments range from 3.0~$M_{\rm Jup}$ to 15~$M_{\rm Jup}$.

The phase of vigourous disk fragmentation ends after $t=0.26$~Myr
and no fragments are seen in the disk afterwards. The disk shrinks notably in size and 
its subsequent evolution is characterized by the presence of a weak flocculent spiral structure. 
This drastic change in the disk evolution as compared to that in model~2 (the latter showing vigorous
fragmentation at these evolutionary times) can be attributed to low angular momentum of the 
infalling external environment, which exerts a negative torque onto the disk. This causes the 
disk to shrink in size. Although the disk mass remains substantial at this phase
(see Section~\ref{diskprop}), a small size ($\la 100$~AU) and high disk temperature ($\approx 200$~K)
make gravitational fragmentation unlikely.

\subsection{The effect of external environment: counterrotating infall}

\begin{figure}
  \resizebox{\hsize}{!}{\includegraphics{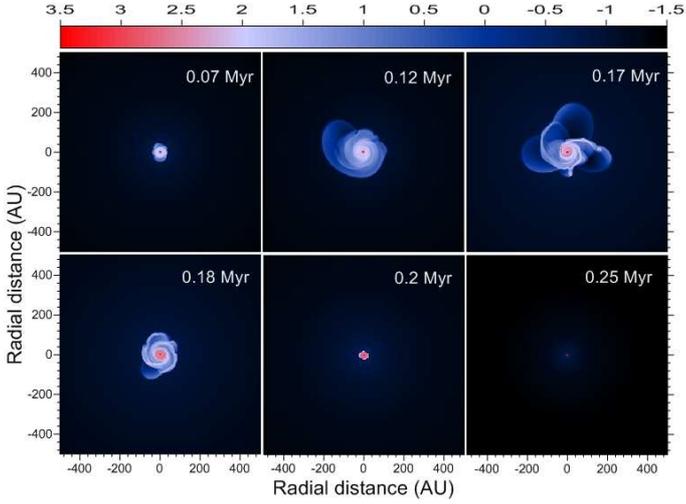}}
  \caption{Similar to Figure~\ref{fig2} but for model~4.}
  \label{fig5}
\end{figure}

In this section we consider the effect of infall from the external environment counterrotating with
respect to the prestellar core. As the first step, we consider model~4 with external environment 
counterrotating (with respect to the core) 
at an angular velocity ten times smaller than that of the core. Figure~\ref{fig5} presents the gas 
surface density maps for model~4 for the same regions and evolution
times as in the isolated model~1. The early evolution of the disk in model~4 ($t\le0.18$~Myr) 
bears some similarity with that in the isolated model~1---the disk rotates 
counterclockwise, it is gravitationally unstable and shows clear signs of fragmentation. 
The fragments have masses around 4.0~$M_{\rm Jup}$.
The drastic change with respect to all models considered so far is seen at $t>0.18$~Myr
when the disk starts quickly shrinking in size and virtually disappears after $t=0.2$~Myr due to a strong
negative torque exerted by the infalling external environment conterrotating with respect to the disk.
The disk looses its rotational support and accretes onto the star, leading to a steep 
increase in the stellar mass as evident in Figure~\ref{fig10}.
This example demonstrates how the counterrotating external environment can significantly shorten 
the lifetime of circumstellar disks.

\begin{figure}
  \resizebox{\hsize}{!}{\includegraphics{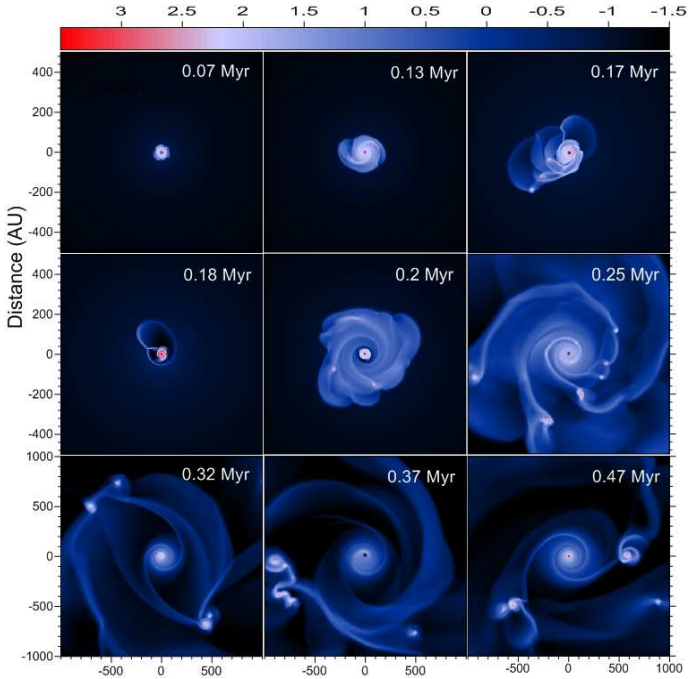}}
  \caption{Gas surface density maps in model~5 in the inner $1000\times1000$~AU box (top row) and 
  $2000\times2000$~AU box (middle and bottom rows) obtained
  at several evolution times (as marked in each panel) after the formation of the central protostar.
  The scale bar is in log g~cm$^{-2}$.}
  \label{fig6}
\end{figure}

As the second step, we consider model~5 with external environment counterrotating (with respect to 
the core) with the same angular velocity at that of the core. Figure~\ref{fig6} presents the 
gas surface density in model~5 in the inner
$1000\times1000$~AU box (top row) and in the inner $2000\times2000$~AU box (middle and bottom rows).
The early evolution ($t\le0.17$~Myr) in models~4 and 5 is very similar -- the disk in both models is
gravitationally unstable and shows signs of fragmentation. The masses of the fragments lie in the
4.0--7.0~$M_{\rm Jup}$ range.

\begin{figure}
  \resizebox{\hsize}{!}{\includegraphics{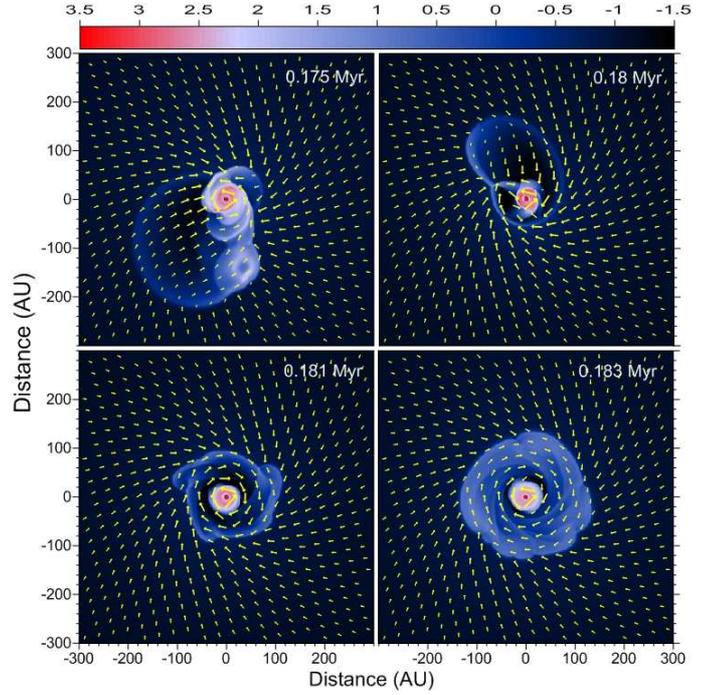}}
  \caption{Development of a cavity in the disk. Shown with yellow arrows is the gas velocity field 
  superimposed on the gas surface density in the inner $600\times600$~AU
  box in model~5. The bar is in log g~cm$^{-2}$.}
  \label{fig7}
\end{figure}

The subsequent evolution of the disk 
in model~5 is however quite different from any model considered so far. First, the disk shrinks in size
to just a few tens AU at $t=0.18$~Myr due to a strong negative torque exerted by the counterotating
infalling material. Then, an outer disk forms rotating in the opposite direction with respect to 
that of the inner, heavily reduced disk. During this transformation, a notable cavity develops
in the gas surface density at a radial distance of the order of 
a few tens of AU. This process is illustrated in Figure~\ref{fig7} showing the gas velocity field
(yellow arrows) superimposed on the gas surface density in the inner $600\times600$~AU box during 
a narrow time period of 8~kyr. The infall of counterrotating envelope onto the disk is already 
evident at $t=0.175$~Myr, leading to a sharp reduction in the disk size by $t=0.18$~Myr. The outer
counterrotating disk starts forming at t=0.181~Myr and becomes fully developed by t=0.183~Myr. 
The inner edge of the cavity is initially located at $r\approx 30$~AU.

\begin{figure}
  \resizebox{\hsize}{!}{\includegraphics{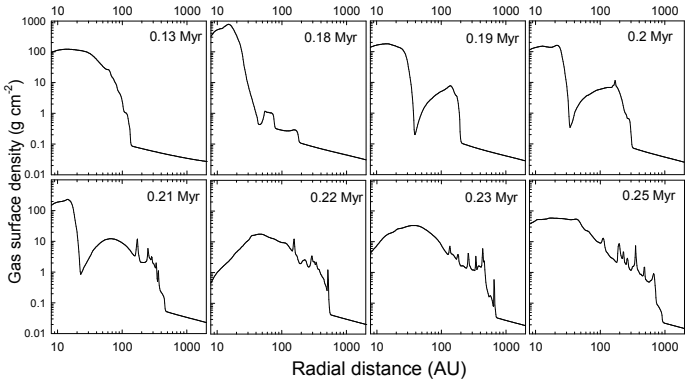}}
  \caption{Time series of azimuthally averaged gas surface density profiles ($\overline{\Sigma}$, solid
  lines) showing the time evolution of the cavity in model~5. }
  \label{fig8}
\end{figure}

The time evolution of the cavity can be best described with a series of azimuthally averaged gas surface
density ($\overline{\Sigma}$)  profiles shown in Figure~\ref{fig8}. The radial profile of $\overline{\Sigma}$
in the early evolution is characterized by a near-constant-density central region ($\la 30$~AU) 
and a steeply declining tail at larger radii. The cavity appears at $t=0.18$~Myr 
and become fully developed by $t=0.19$~Myr, with the maximum contrast in $\overline{\Sigma}$ between
the lowest value in the cavity and the maximum values in the inner and outer disks on the order 
of 900 and 40, respectively. As time progresses, the cavity moves
closer to the central star because the matter from the outer disk flows through the cavity onto the
inner disk, exerting a negative torque on to the latter.  By $t=0.22$~Myr the cavity turns into 
a central hole when the inner disk becomes smaller than the size of our sink cell.  
Finally, by $t=0.25$~Myr the central hole fills in and the subsequent profile
of $\overline{\Sigma}$ becomes typical for massive, strongly unstable disks with local sharp maxima
representing fragments embedded in the disk.

After t=0.2~Myr the disk
experiences a second vigorous episode of disk fragmentation leading in the end to its almost complete
disintegration and formation of a multiple system of fragments connected with dense filaments. 
The masses of the fragments at this stage reach a much higher value, ranging from 7.0 to 82 Jupiter
masses.

\begin{figure}
  \resizebox{\hsize}{!}{\includegraphics{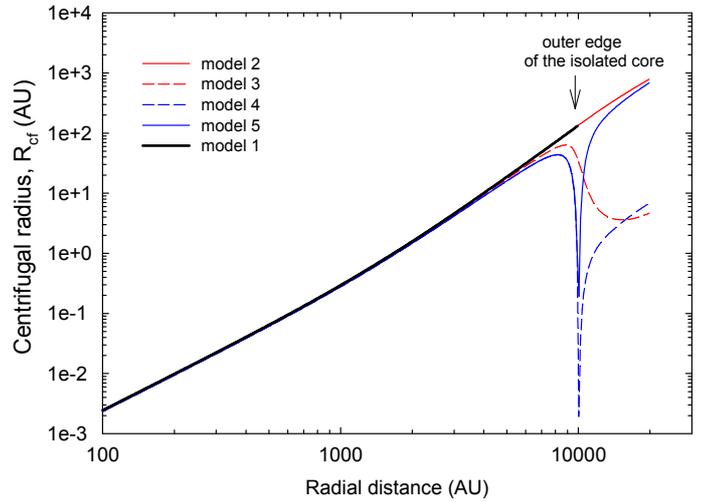}}
  \caption{Centrifugal radius $R_{\rm cf}$ as a function of radial distance in models 1--5.}
  \label{fig9}
\end{figure}

\begin{figure}
  \centering
  \vskip 3.0 cm
  \resizebox{\hsize}{!}{\includegraphics{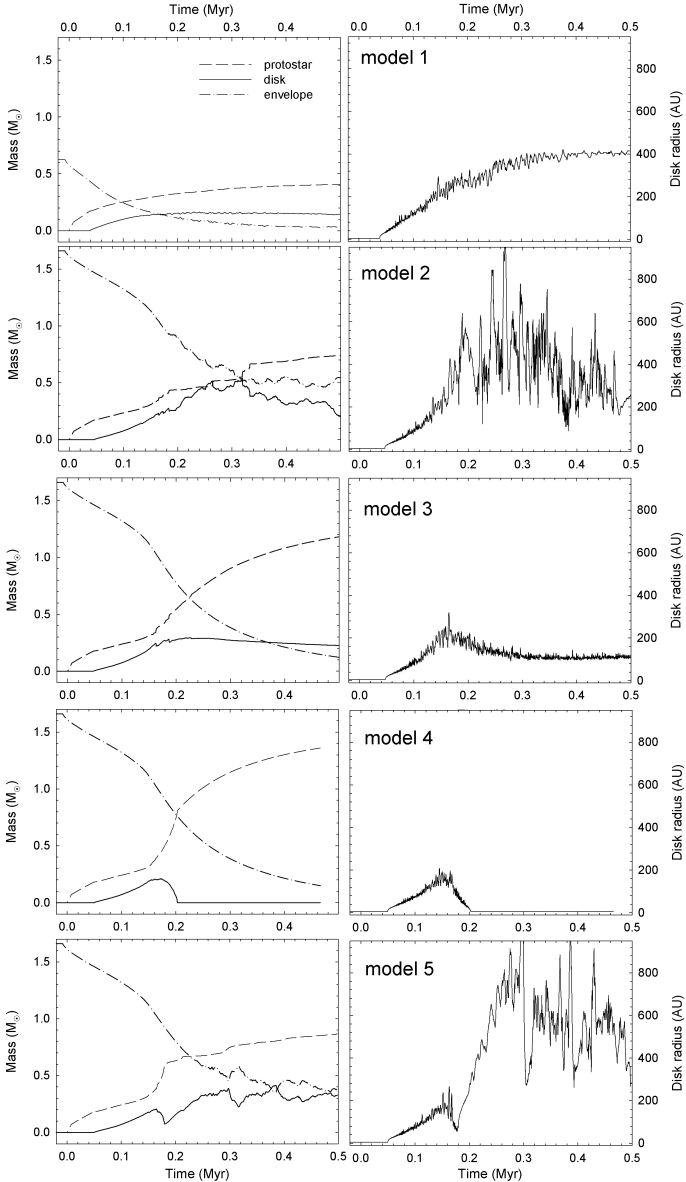}}
  \caption{Left column: masses of the disk (solid lines), star (dashed lines), and envelope (dash-dotted
  lines) as a function of time elapsed since the formation of the protostar in models~1--5 (from top
  to bottom rows).  Right column: the corresponding disk radii.  }
  \label{fig10}
\end{figure}

The difference in the properties of circumstellar disks formed in the considered models 
can be understood by analyzing the centrifugal radius of matter initially located 
at a distance $r$ from the star:
\begin{equation}
R_{\rm cf}= {J^2(r) \over G M(r)},
\end{equation} 
where $J(r)=r^2 \Omega$ is the specific angular momentum at a radial distance $r$,
$G$ is the gravitational constant, and $M(r)$ it the mass enclosed within distance $r$. 
Figure~\ref{fig9} presents the centrifugal radius as a function of distance for various initial distributions
of mass and angular momentum in models~1--5.
More specifically, the thick black line shows $R_{\rm cf}$ for the isolated pre-stellar core 
in model~1, while the solid red and dashed red lines provide the centrifugal radius
in models~2 and 3 characterized by a corotating external environment. The solid blue and 
dashed blue lines
show $R_{\rm cf}$ for models~4 and 5 characterized by a counterrotating external environment.

The centrifugal radius in model~1 is gradually increasing with distance and reaches a maximum value
of 132~AU for the material located at the core outer edge (shown by the arrow). We note however 
that the disk in model~1 has in fact grown to a size greater than suggested by the maximum centrifugal
radius of $R_{\rm cf}=132$~AU (see Fig.~\ref{fig2}) due to viscous spreading. 

The radial profile of $R_{\rm cf}$ in model~2 (solid red line) is similar to that 
of model~1, except that the matter of the external
environment is characterized by a gradually increasing $R_{\rm cf}$ reaching a maximum 
value of 791~AU. As a net result, the disk in model~2 grows 
to a larger size and becomes more massive than in model~1 (see Fig.~\ref{fig10}), breaking 
finally into a series of fragments due to violent gravitational instability. 
In model~3, however, the external environment has a much smaller angular velocity than that of 
the core. As a consequence, the centrifugal radius of the external environment gradually declines 
to just several AU (dashed red line). The infalling material from the 
external environment with a substentially sub-Keplerian rotation 
exerts an ever increasing negative torque onto the disk outer regions. 
This causes the disk to shrink in size with time as is evident in Figures~\ref{fig4} and \ref{fig10}.

The last two models~4 and 5 are characterized by external environment rotating in the opposite direction
to that of the core.  In model~4, characterized by angular velocity of the external environment 10 times
smaller than that of the core, the centrifugal radius
first grows with distance (as in model~1) but then shows a sharp drop near the core outer edge 
where the angular velocity changes its sign. $R_{\rm cf}$ of the external material 
always stays below several AU. As a result, 
the infalling external material with a counterrotating sub-Keplerian velocity exerts a strong 
negative torque onto the disk outer regions. The disk shrinks to a size smaller than 
the size of our sink cell (5~AU) and virtually disappears in Figure~\ref{fig5}.

In model~5 the angular velocity of the counterrotating external environment
is the same as that of the core. As a consequence, $R_{\rm  cf}$ first drops to a small value 
near the core outer edge (where the rotation changes its sign) but then quickly grows to 
a maximum value of 685~AU. This causes the disk first to shrink in size,
when the low-$R_{\rm cf}$ material from the external environment  hits the disk and 
starts extracting disk's angular momentum. Then, the infalling external material characterized by 
an ever growing $R_{\rm cf}$ hits the centrifugal barrier 
just outside the heavily reduced inner disk and the outer, counterrotating disks begins to form.
The transition region between the inner and outer disks, where rotation changes its direction and
the matter lacks centrifugal support, is manifested by a density gap clearly visible
at $t=0.2$~Myr in Figure~\ref{fig6}.

\section{Comparison of disk properties}
\label{diskprop}
In this section, we compare various properties of our model disks including the disk masses 
and sizes, which were computed using the disk tracking algorithm. 
The disk mass and radius are calculated
at each time step by disentangling the disk and infalling envelope
on the computational mesh. We do this in practice by adopting
a surface density threshold of $\Sigma_{\rm cr}$=0.5~g~cm$^{-2}$ between the disk and
envelope and also using the radial gas velocity profile. This method
is described in detail by \citet[][see fig. 1 and the pertaining text]{Vor2011}.
The values of the disk radius and mass both
depend somewhat on the adopted threshold. Nevertheless, a lower value
of $\Sigma_{\rm cr}$=0.1~g~cm$^{-2}$ results only in a few percent increase
in the net disk mass \citep[see Fig. 2 in][]{Dunham2014}.

We found that our algorithm generally works well, but may somewhat overestimate the disk mass in
the early Class 0 stage (when the inner part of the infalling envelope may
have densities exceeding $\Sigma_{\rm cr}$) and somewhat underestimate the disk
mass in the late Class I and Class II stages (when the disk spreads out and its density
drops below $\Sigma_{\rm cr}$ in the outer parts). 
The envelope mass includes both the matter in the core and external environment (when present).

The left column in Figure~\ref{fig10} presents the integrated disk masses $M_{\rm d}$ 
(solid lines), stellar masses $M_\ast$ (dashed lines) and envelope masses $M_{\rm env}$ 
(dash-dotted lines) as a function of time passed since the formation
of the central protostar, while the right column shows the corresponding disk radii $R_{\rm d}$. 
Each row corresponds to a particular model as indicated in the figure. 
In model~1, the disk mass steadily grows and reaches a maximum value of 
$M_{\rm d}\approx 0.16~M_\odot$ at $t\approx 0.2$~Myr, gradually declining afterwards. 
The growth rate of  the protostar exceeds that of the disk and the final mass of the star 
is expected to be about $M_\ast=0.45~M_\odot$. The envelope mass steadily decreases and the 
model enters the class II stage, defined as the time when less than 10\% of the initial 
mass reservoir is left in the envelope, at $t\approx0.25$~Myr. The disk grows in radius showing small
radial pulsations and reaches a maximum value of $R_{\rm d}\approx 400$~AU at $t\approx0.5$~Myr.
All in all, this model demonstrates a standard behaviour as has been found in previous numerical
simulations of isolated collapsing cores \citep[e.g.][]{Vor2011}.

The properties of the forming star plus disk system in model~2 are strikingly 
different from those of model~1. First, the disk mass reaches a maximum value of $0.54~M_\odot$, 
three times greater than that in model~1. Second, the disk radius exhibits large variations, which are
indicative of significant radial motions within the disk caused by the gravitational interaction 
between the fragments and elements of spiral arms\footnote{These numbers should be taken with 
caution because our disk tracking algorithm may be inaccurate on systems with violently 
fragmenting disks.}.
Third, the central star experiences several episodes of sharp increase in mass
caused by accretion of infalling fragments, the effect discussed in more details in Section~\ref{accrete}
below. The final stellar mass is expected to be around $0.8~M_\odot$. 
After $t=0.5$~Myr, there is still 30\% of the total mass available for accretion onto the disk+star
system, indicating that the system is still in the embedded phase. We note that
the mean duration of the embedded phase obtained in numerical hydrodynamics simulations of 
collapsing isolated cores ($\approx 0.15$~Myr, albeit with a large scatter depending on the core mass)
appears to be shorter by a factor of 3--4 than that obtained by analyzing the
number of embedded and Class II sources in star-forming regions \citep{Vorobyov2010,Dunham2012}.
A prolonged accretion from the external environment may provide a possible solution to this problem.

Model~3 with a slowly corotating external environment is peculiar in several aspects.
On the one hand, the disk radius stays around 
100~AU during most of the evolution, exceeding 200~AU only for a limited period. 
On the other hand, the disk mass becomes at least a factor of 2 greater than that 
in model~1, indicating the formation of a dense and compact disk.
Due to an elevated mass transport through the disk caused by infall of low-angular-momentum
material from the external environment, the final stellar mass is likely 
to exceed $M_\ast=1.2~M_\odot$, forming in the long run 
a substantially more massive object than in models~1 and 2. The envelope mass after 
$t=0.5$~Myr is 5\% of the total available mass, indicating that the system has entered 
the Class II stage. To summarize, model~3 is capable of forming compact and dense disks 
around solar-mass stars.

Model~4 with a slowly counterrotating external environment is unique in its kind. 
The negative torque exerted onto the disk by infalling counterotating material leads
to complete accretion of the disk onto the protostar by $t=0.2$~Myr. A small 
circumstellar disk may still exist on sub-AU scales, as suggested by the centrifugal radius 
in Figure~\ref{fig9}, but we cannot confirm its presence in our numerical simulations 
due to the use of a sink cell at $r=5$~AU. In the subsequent evolution, the forming star 
is likely to accrete material either from a small sub-AU
disk or directly from the infalling envelope. The fraction of stars with detectable near-infrared 
disk emission seems to approach 80-90\% for objects with an age smaller than 1.0~Myr \citep{Hernandez2007},
suggesting that systems like those explored in model~4 are rare.

Finally, model~5 presents a very curious case showing the reversal of the disk rotation due to
the infall of counterrotating external material. The initial disk growth 
is followed by a sharp decline when the disk mass and size drop to just $0.07~M_\odot$ 
and $\la 100$~AU at $t\approx 0.18$~Myr, manifesting the beginning of the disk reversal phase. 
In the subsequent stage, the disk evolution is similar to that in model~2 showing
vigorous gravitational instability and fragmentation.  The final stellar mass is expected to be slightly
below that of the Sun. The formation of systems with disks counterrotating to that of the star is 
rare, with RW Aur A being one possible example showing both the disk and jet 
counterrotating with respect to the rotation of the host binary system \citep{Woitas2005,Bisikalo2012}.

We want to mention that the gas temperature in models~3--5, which feature compact disks during the evolution,
becomes comparable to or higher than 200~K at $r$=10-20~AU, a value which is at least a factor 
of two greater than that in the isolated model~1. We cannot calculate the gas temperature 
at the sub-AU scales, but by extrapolating the $T\propto r^{-1/2}$ profile we speculate that 
the temperature at $r$=0.5~AU can reach values on the order of 1200~K, which is close to 
what is needed for the formation of calcium-aluminium-rich inclusions (CAIs). Numerical 
simulations with a smaller sink cell are planned for the future to investigate this possibility.

Table~\ref{table2} summarizes the main properties of the forming objects in models~1-5. The mean values
are calculated starting from the formation of the disk and ending at $t=0.5$~Myr after the 
formation of the protostar. In model~4, the mean values were taken over the time period covering 
the actual existence of the disk.

\begin{table}
\center
\caption{Main model properties}
\label{table2}
\begin{tabular}{cccc}
\hline\hline
Model & Final stellar & Mean/max disk & Mean/max disk   \\
 & mass $(M_\odot)$ & mass ($M_\odot$) & radius (AU) \\
\hline
1 & 0.42 & 0.13/0.16 & 295/426   \\
2 & 0.74 & 0.29/0.54 & 319/1057   \\
3 & 1.23 & 0.22/0.29 & 122/318   \\
4 & 1.3  & 0.1/0.21 & 83/210   \\
5 & 0.9 & 0.16/0.41 & 286/1370  \\
\hline
\end{tabular}
\end{table}

\section{Accretion and infall rates}
\label{accrete}

\begin{figure}
  \centering
  \resizebox{\hsize}{!}{\includegraphics{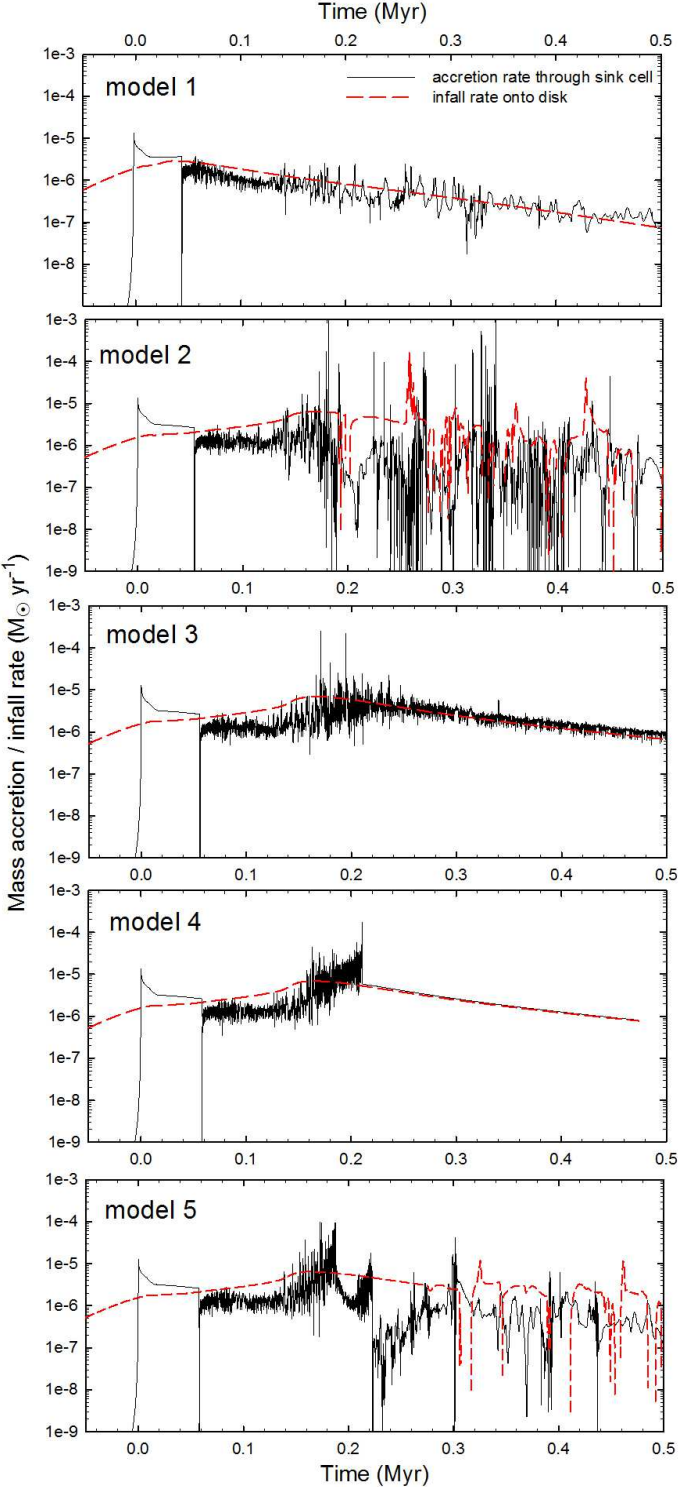}}
  \caption{Mass accretion rates through the sink cell at $r=5$~AU (black solid lines) 
  and mass infall rates at $r=2000$~AU (red dashed lines) in models~1--5 (from top to bottom).}
  \label{fig11}
\end{figure}

In this section we compare the mass accretion rates through the sink cell $\dot{M}$ 
and mass infall rates at a distance of 2000~AU from the star $\dot{M}_{\rm infall}$ in our five 
models. Figure~\ref{fig11} presents $\dot{M}$ (solid lines) and $\dot{M}_{\rm infall}$ (red dashed 
lines) as a function of time elapsed since the formation of the first hydrostatic core (FHC) 
in models~1--5 (from top to bottom).

The accretion rate in model~1 shows a time behavior typical for weakly gravitationally unstable
disks \citep{VB10}. $\dot{M}$ is negligible in the pre-stellar phase and rises
to a maximum value of $1.3 \times 10^{-5}$~yr$^{-1}$ when the FHC
forms at t = 0 Myr. The subsequent short period of evolution is characterized by a
gradually declining $\dot{M}$ when the material from the core lands
directly onto the forming star. A sharp drop in the accretion rate at $t = 0.045$~Myr 
manifests the beginning of the disk formation
phase when the infalling core material hits the centrifugal
barrier near the sink cell and the accretion rate drops to a negligible value.
However, the process of mass loading from the core continues,
the disk grows in mass and size, and the disk quickly becomes gravitationally unstable. 
In the subsequent evolution $\dot{M}$ shows order-of-magnitude flickering and gradually declines with time to a value of $10^{-7}~M_\odot$~yr$^{-1}$ at $t=0.5$~Myr.
We note that model~1 exhibits no accretion bursts with magnitude $\ga 10^{-4}~M_\odot$~yr$^{-1}$, 
as might be expected for  models with  fragmenting disks whereby fragments migrate onto the 
star due to gravitational interaction
with spiral arms \citep{VB06,VB10}. This indicates that the fragments in model~1 (see Fig.~\ref{fig2})
were tidally destroyed before reaching the inner sink cell.
We also note that $\dot{M}_{\rm infall}$ in model~1 exceeds $\dot{M}$ only during
the early evolution of the disk ($t\approx0.05-0.2$~Myr) and both values are similar
in magnitude  in the subsequent evolution. The inequality $\dot{M}_{\rm infall}>\dot{M}$ is one of 
the conditions for disk gravitational fragmentation \citep[e.g.][]{VB10,Kratter2010},
explaining the presence of fragments in the disk of model~1.

The early evolution of $\dot{M}$ in model~2 is similar to that of model~1.
A notable difference comes about after $t=0.15$~Myr when $\dot{M}$
begins to show variations by several orders of magnitude.  Accretion bursts, some of which 
exceeding in magnitude $10^{-4}~M_\odot$~yr$^{-1}$, are caused by disk gravitational 
fragmentation followed by migration of the fragments 
onto the forming protostar -- a phenomenon known as the burst mode
of accretion \citep{VB06,VB10}. The number of bursts in this particular model amounts to a 
few tens and the burst activity subsides in the early Class II stage. 
The mass infall rate $\dot{M}_{\rm infall}$
is greater on average than $\dot{M}$, thus leading to the accumulation of mass in the disk and 
strengthening its gravitational instability and propensity to fragment.   
To check this supposition, we calculated the spatial distribution of the Gammie parameter 
${\cal G}=t_{\rm c} \Omega$, where $t_{\rm c}=e/\Lambda$ is the local cooling time, and 
the Toomre parameter $Q=c_{\rm s} \Omega/(\pi G \Sigma)$. 
Disk fragmentation is supposed to occur in regions where both quantities become
smaller than unity \citep[e.g.][]{Gammie2001,Rice2003}. 
We found that during the initial disk evolution, unaffected by infall
from the external environment, the $Q$-parameter becomes smaller than unity only in a few spots in 
the disk, while the regions with ${\cal G}<1.0$ cover a much larger area, indicating that the
insufficient strength of gravitational instability (rather than slow disk cooling) 
is the main reason hindering fragmentation. On the other
hand, during the external infall stage, the regions with $Q<1$ occupy a notably larger area, while 
the spatial area with ${\cal G}<1$ remains similar to what was found during the early stage, meaning
that the infall from the external environment is the dominant effect driving the disk to the fragmentation
boundary.

We note that $\dot{M}_{\rm infall}$ sometimes
shows large variations caused by the perturbing influence of the disk. More specifically, 
the gravitational interaction between fragments and spiral arms within the disk 
sometimes leads to ejections of individual fragments or elements of spiral arms to distances on the
order of thousands AU. 

The mass accretion rate in model~3 with a slowly corotating external environment exhibits
a much smaller accretion variability than that of model~2. There are only two strong accretion
bursts around $0.18-0.2$~Myr and the later evolution of $\dot{M}$ is characterized 
only by small-scale flickering. On average, $\dot{M}$ is higher in model~3 than in models~1 and 2, 
staying during the initial 0.5~Myr of evolution in the $10^{-6}-10^{-5}~M_\odot$~yr$^{-1}$ range. 
The matter from the external environment falling in onto the disk at a substantially 
sub-Keplerian speed acts to reduce the disk rotational support, thus increasing 
the mass flow through the disk onto the star.
We note that $\dot{M}_{\rm infall}$ is greater than $\dot{M}$ at $t<0.20$~Myr, causing disk
fragmentation in the early disk evolution (see Fig.~\ref{fig4}), but both values become comparable 
in the later evolution, indicating the presence of efficient
mass transport through the compact and hot disk due to the increasing strength of viscous torques.

The effect of the external environment in model~4 is manifested
by a gradual increase of the mass accretion rate after $t=0.16$~Myr to 
$\ga 10^{-5}~M_\odot$~yr$^{-1}$ (on average).  This increase is even more 
pronounced than in model~3 due to the fact that the infalling external material in model~4 
is counterrotating with respect to the disk, thus exerting a stronger negative torque onto the disk
outer regions than in model~3. As a net result, the disk loses rotational support and 
the mass transport
rate through the disk increases leading to a complete disappearance of the disk after 
$t=0.22$~Myr. In the subsequent evolution, both $\dot{M}$ and $\dot{M}_{\rm infall}$ are 
nearly identical and slowly decline with time.

Finally, the early evolution of $\dot{M}$ in model~5 is very similar to that in model~4. After
an initial period of near constant accretion at a value of $\approx 10^{-6}~M_\odot$~yr$^{-1}$, 
the mass accretion rate shows a transient increase by an order of magnitude during $t=0.15-0.22$~Myr.
During this period, the matter is accreted from the inner disk, which is gradually losing 
its rotational support and shrinking in size due to the infall of matter from the counterrotating 
external environment.
At $t\approx0.23$~Myr, the inner disk reduces to a size smaller than the sink cell and $\dot{M}$
drops substantially. In the subsequent evolution, the matter is accreted from the outer, counterrotating
disk. Because the infall rate  $\dot{M}_{\rm infall}$ is greater on average than the mass 
accretion rate onto the star $\dot{M}$, the outer disk quickly becomes gravitationally unstable, 
leading high-amplitude variations in $\dot{M}$.
It is interesting to note that the later evolution shows only one moderate accretion burst at
$t\approx 0.3$~Myr, implying that fragments forming in the disk (see Fig.~\ref{fig6}) are either
tidally dispersed or may stay in the system with a possibility of forming in the long run a 
multi-component stellar/planetary system.

\section{Discussion: the variety of disk properties}
Observations of star-forming regions indicate that the fraction of (sub-)solar-mass stars 
with circumstellar disks approaches 80\%-90\% for objects with an age equal to or smaller 
than 1.0~Myr \citep{Hernandez2007}. The young disks seem to
have diverse properties with masses ranging from a few Jupiters to a fraction of solar mass 
and sizes from a few to hundreds AU \citep{Andrews09,Jorgensen09,Eisner2012,Tobin2012}.

The observed variety of disk properties can in principle be reproduced by variations in the mass, 
angular momentum, and magnetic field strength of {\it isolated} parental cores. For instance, 
numerical hydrodynamics simulations of the gravitational collapse of non-magnetized isolated cores 
\citep{Vor2011} yield  the following near-linear relation between the disk masses and the masses of the central sub-stellar/low-mass object 
\begin{eqnarray}
M_{\rm d,CO}&=&\left( 0.73^{+0.11}_{-0.09} \right)\, M_{\ast,\rm CO}^{1.05\pm 0.07}, \\
M_{\rm d,CI}&=&\left( 0.65^{+0.04}_{-0.05} \right)\, M_{\ast,\rm CO}^{1.0\pm 0.04},
\end{eqnarray}
where the indices CO and CI correspond to the class 0 and class I phase, respectively. The minimum
and maximum disk masses span a range from $0.02~M_\odot$ to $0.4~M_\odot$. 
At the same time, observations of embedded sources infer disks with the lower limit extending
to just one Jupiter mass \citep{Jorgensen09,Tobin2012} and, in some cases, fail to detect 
disks at all \citep{Maury2010}.
Systems with very-low-mass disks ($M_{\rm d}\la 0.01~M_\odot$) can be the result of efficient 
magnetic braking operating in the embedded phase of star formation 
\citep{Machida2011b,Seifried2012} or can simply be formed from cores with very low 
angular momentum and, subsequently, having a very small centrifugal radius.
The situation is similar for disk radii -- numerical simulations of isolated non-magnetized cores 
\citep{Vor2011} seem to yield disk sizes that are a factor of several greater than inferred 
from observations \citep{Tobin2012,Eisner2012}. 

In this paper, we have explored the evolution of pre-stellar cores submerged into an external 
environment and found that its effect can add another dimension to the variety 
of disk properties around (sub-)solar-mass stars. Depending on the value of angular velocity 
and the direction of rotation of the
external environment with respect to the core the resulting disks can be characterized by vastly different
physical parameters such as masses, sizes, lifetimes, and accretion 
rates onto the host star. The external environment can affect the strength of gravitational 
instability in the disk and hence the likelihood of giant planet and brown dwarf formation. 
The effect can actually work both ways: suppressing or promoting disk gravitational
fragmentation depending on the amount of rotation of the external material.

Models~3 and 4 with a slowly rotating external environment predict the formation of 
compact disks with radii $\le 200$~AU and, in some cases, even smaller than
5~AU (the size of our central sink cell). Even model~2 with a fast rotating environment
is characterized by a more compact disk (than the isolated model~1) in the late evolution, 
which  is likely explained by the loss of angular momentum carried away by fragments accreted onto the
star\footnote{We have not calculated the angular momentum flux through the sink cell. This study will
be presented in a follow-up paper.}.
Observational estimates of disk radii seem to 
indicate that the majority of embedded and T~Tauri disks  are indeed rather compact, 
$R_{\rm d}=50-200$~AU \citep{Vicente2005,Tobin2012,Murillo2013}. 

Perhaps, the most interesting finding of our numerical modelling is the ability to form disks 
with cavities or holes in the
case of counterrotating external environment. These features are transient with the lifetime 
of a few tens kyr and are characterized by conterrotating inner and outer disks. In our model,
the gap is located at $\approx 30-50$~AU, which is similar with the position of
the gap in, e.g., LkH$\alpha$~330 \citep{Brown2008}. The gap migrates inward with time gradually
transforming into a central hole. Though being a short-lived phenomenon, this mechanism of 
the gap/hole formation can present an interesting alternative to other mechanisms such as 
disk photoevaporation by the central star  \citep{Clarke2001,Owen2010}
or gravitational clearing due to a massive planet \citep[e.g.][]{Kley2001}.

After the inner, corotating disk vanishes due to accretion of its material onto the star, 
a stellar system with a disk counterotating with respect to that of the star emerges.
If the disk is sufficiently massive to fragment, this process may lead to the
formation of a system with 
sub-stellar/very-low-mass components counterrotating with respect 
to the host star. This phenomenon may, however, be transient in nature.
The matter is that angular momentum in the external environment is usually much greater 
than that of the star and the latter may change stellar rotation to match that of the infalling material.
Although the efficiency of this process should depend on how much angular momentum is lost via jets/outflows
and whether or not a binary/multiple stellar system forms instead of a single star,
such counterotating systems are nevertheless expected to be rare, with RW Aur A being one possible 
example \citep{Woitas2005,Bisikalo2012}.

We note that we have considered a rather simplified initial setup, treating the external 
environment as a homogeneous medium of constant density and angular velocity. 
In reality, however, the external environment
is likely to have a complicated velocity and density structure as often seen in
numerical hydrodynamics simulations of clustered star formation.
For instance \citet{Bate2010}, \citet{Fielding2014}, and \citet{Padoan2014} 
demonstrated that individual cores may accrete mass and angular momentum
at a highly non-steady rate due to a filamentary and chaotic structure of the intracluster 
medium, leading to the formation of misaligned star-disk systems 
(the misalignment in some transient cases exceeds $90^\circ$).
It is therefore possible that the angular momentum vector of the accreted material can 
undergo significant changes, including its orientation with respect to that of the core.
Moreover, the dynamics of dust particles needs to be considered as well because 
the holes/gaps in the disk are often identified through the lack of dust emission there.
Finally, we acknowledge that the external mass in our models ($M_{\rm ext}=1.0~M_\odot$) 
is greater than that of the core (0.63~$M_\odot$), which may have amplified the effect 
of infall from the external environment. Models with a smaller $M_{\rm ext}$ need to 
be also investigated in order to evaluate the dependence of our results on the 
available mass in the external reservoir.
We plan to consider
more realistic initial configurations, smaller external mass, and the dynamics of 
dust in a subsequent study.

\section{Conclusions}
In this paper we have performed numerical hydrodynamics simulations of the gravitational collapse
of a pre-stellar core with sub-solar mass submerged into a low-density external environment 
characterized by different magnitude and direction of rotation with respect to the core. 
We followed the evolution of our models through the star and 
disk formation stage and terminated the simulations when the age of the forming star started to exceed
0.5~Myr. In all models, the mass of the core and external environment were fixed at $0.63~M_\odot$
and $1.0~M_\odot$, respectively, in order to exclude systematic differences due to variations
in these parameters. We found that depending on the value of angular velocity and 
the direction of rotation of the external environment with respect to the core the resulting disks 
can be characterized by vastly different physical properties. 

Our most interesting finding is that infall of material from the external environment counterrotating
with respect to the core may lead to the formation of the inner and outer disks rotating in 
the opposite directions to each other. The transitional region between the disks manifests itself
as a deep gap in the gas density, which migrates inward due to accretion of the inner disk onto the
protostar. After about several tens of kyr, the gap turns into a central hole, which fills 
in after another few tens of kyr. The formation of a transient gap/hole due to infall 
from counterrotating external environment presents an interesting alternative to other mechanisms 
of gap/hole formation such as disk photoevaporation \citep{Clarke2001,Owen2010}
or gravitational clearing due to a massive planet \citep[e.g.][]{Kley2001}.
Our main results can be summarized as follows.
\begin{itemize}
\item Pre-stellar cores submerged into an external environment corotating at an  
angular velocity similar in magnitude to that of the core tend to form extended and
strongly unstable disks,  
which fragment into several objects connected by dense filaments and  may form 
in the long run systems with sub-stellar/very-low-mass companions. 
The mass accretion rate onto the primary shows a highly variable character with 
episodic bursts caused by the infall of some of the fragments onto the star.
\item Pre-stellar cores embedded into a slowly corotating external environment (with angular velocity
ten times smaller than that of the core) are likely to form compact ($R_{\rm d}\le 200$~AU)
disks, stable against gravitational 
fragmentation due to their small size. These disks can drive an order of magnitude higher 
mass accretion rates onto the star than those formed from isolated cores.
\item Pre-stellar cores submerged into a slowly {\it counterrotating} external environment are expected
to form compact ($R_{\rm d}\le 200$~AU) and short-lived ($\la \mathrm{a~few}~\times 10^{5}$~yr) disks
due to strong negative torques exerted by the infalling external material.
\item Disks formed from pre-stellar cores embedded into an external environment {\it counterotating} with respect to the core at a similar angular velocity are likely to change their initial direction
of rotation. In the process of this transformation, the inner and outer disks form
counterotating with respect to each other and separated by a transient gap in the gas density. 
In the long run, the inner disk accretes onto the star and a transient stellar system with 
sub-stellar/very-low-mass components counterrotating with respect to that of the star may emerge
due to gravitational fragmentation of the outer, strongly unstable disk. 
\end{itemize} 

We thank the referee for constructive comments that helped to improve the manuscript.
This project was partly supported by the Russian Ministry of Education and Science Grant 
(state assignment) 3.961.2014/K. MG acknowledges support by the Austrian FWF through the NFN 
project grant S116 "Pathways to Habitability: From Disks to Active Stars, Planets and Life", 
and the related subproject S116 604-N16 "Radiation \& Wind Evolution from T Tauri Phase to ZAMS and Beyond". The simulations were performed
on the Shared Hierarchical Academic Research Computing Network
(SHARCNET), on the Atlantic Computational Excellence Network (ACEnet),
and on the Vienna Scientific Cluster (VSC-2). This publication is supported by the 
Austrian Science Fund (FWF).

\end{document}